%% file: milliQan_LOI.tex
\documentclass[superscriptaddress,preprint,preprintnumbers,amsmath,amssymb]{revtex4}

\usepackage{graphicx}
\usepackage{dcolumn}
\usepackage{bm}
\usepackage{subfig}

\usepackage{verbatim}
\usepackage{amsfonts}
\usepackage{amssymb}
\usepackage{amsmath}
\usepackage{makeidx}
\usepackage{color}
\usepackage{latexsym}
\usepackage{slashed}
\usepackage[all]{xy}
\usepackage{hyperref}  
\newcommand{\be}{\begin{eqnarray}}
\newcommand{\ee}{\end{eqnarray}}

\newcommand{\GeV}{~\mathrm{GeV}}

\newcommand{\MQ}{M_{\rm mCP}}

\newcommand{\gsim}{\lower.7ex\hbox{$\;\stackrel{\textstyle>}{\sim}\;$}}
\newcommand{\lsim}{\lower.7ex\hbox{$\;\stackrel{\textstyle<}{\sim}\;$}}
\newcommand{\ZZ}{{\rm Z}^0}

\newcommand{\thetaW}{\theta_{_W}}

\begin{document}


\title{A Letter of Intent to Install a Milli-charged Particle Detector at LHC P5} 

\author{Austin Ball} 
\affiliation{CERN}

\author{Jim Brooke}
\affiliation{University of Bristol}

\author{Claudio Campagnari}
\affiliation{University of California, Santa Barbara}

\author {Albert De Roeck}
\affiliation{CERN}

\author{Brian Francis}
\affiliation{The Ohio State University}

\author{Martin Gastal}
\affiliation{CERN}

\author{Frank Golf}
\affiliation{University of California, Santa Barbara}

\author{Joel Goldstein}
\affiliation{University of Bristol}

\author{Andy Haas}
\affiliation{New York University}

\author{Christopher S. Hill}
\affiliation{The Ohio State University}

\author{Eder Izaguirre}
\affiliation{Perimeter Institute for Theoretical Physics}

\author{Benjamin Kaplan}
\affiliation{New York University}

\author{Gabriel Magill} 
\affiliation{McMaster University}
\affiliation{Perimeter Institute for Theoretical Physics}

\author{Bennett Marsh}
\affiliation{University of California, Santa Barbara}

\author{David Miller} 
\affiliation{University of Chicago}

\author{Theo Prins}
\affiliation{CERN}

\author{Harry Shakeshaft}
\affiliation{CERN}

\author{David Stuart}
\affiliation{University of California, Santa Barbara}

\author{Max Swiatlowski}
\affiliation{University of Chicago}

\author{Itay Yavin}
\affiliation{McMaster University}
\affiliation{Perimeter Institute for Theoretical Physics}

%

\date{\today}

\begin{abstract}
In this LOI we propose a dedicated experiment that would detect ``milli-charged" particles produced by pp collisions at LHC Point 5. The experiment would be installed during LS2 in the vestigial drainage gallery above UXC and would not interfere with CMS operations. With 300~fb$^{-1}$ of integrated luminosity, sensitivity to a particle with charge $\mathcal{O}(10^{-3})~e$ can be achieved for masses of $\mathcal{O}(1)$~GeV, and charge $\mathcal{O}(10^{-2})~e$ for masses of $\mathcal{O}(10)$~GeV, greatly extending the parameter space explored for particles with small charge and masses above 100 MeV. 
\end{abstract}

\maketitle

\section{Introduction \label{sec:Intro}}
\input{intro}

\section{Site Selection \label{sec:site}}
\input{site.tex}

\section{Relationship with CMS \label{sec:cms}}
\input{cms.tex}

\section{Detector Concept \label{sec:det}}
\input{det.tex}

\section{Mechanics, Cooling, and Magnetic Shielding \label{sec:infra}}
\input{infra.tex}

\section{Power and Calibrations \label{sec:pow}}
\input{calib.tex}

\section{Trigger and Readout \label{sec:daq}}
\input{readout.tex}

\section{Backgrounds \label{sec:bkg}}
\input{bkg.tex}

\section{Simulations and Sensitivity \label{sec:sens}}
\input{sim.tex}

\section{Timeline and Next Steps \label{sec:timeline}}
\noindent We aim to have the experiment ready for physics during Run 3. To that end, we envisage the following timeline:

\begin{itemize}
\item Construct small fraction of detector ($\sim10\%$) in next 2 yrs  
\item Install partial detector in PX56 by end of Run 2 (YETS 2017 + TS in 2018)
\item Commission and take data in order to evaluate beam-on backgrounds {\it in situ}
\item Construction + Installation of remainder of detector during LS2 (2019--2020)
\item Final commissioning by spring 2021
\item Operate detector for physics for duration of Run 3 and HL-LHC (mid 2021--)
\end{itemize}

\noindent The next step in the milliQan project is to seek external funding to enable at least the 10\% construction. No such funding has yet been secured for this project, but one or more proposals to one or more funding agencies are being prepared for the near future. 

\section{Summary \label{sec:end}}
\noindent In this LOI we have proposed a dedicated experiment that would detect ``milli-charged" particles produced by pp collisions at LHC Point 5. The experiment would be installed during LS2 in the vestigial drainage gallery above UXC and would not interfere with CMS operations. Our calculations and simulations indicate that with 300~fb$^{-1}$ of integrated luminosity, sensitivity to a particle with charge $\mathcal{O}(10^{-3})~e$ can be achieved for masses of $\mathcal{O}(1)$~GeV, and charge $\mathcal{O}(10^{-2})~e$ for masses of $\mathcal{O}(10)$~GeV. This would greatly extend the parameter space explored for particles with small charge and masses above 100 MeV. We have performed sufficient R\&D to encourage us to proceed with securing funding for the project, and with this letter of intent we express the intention to do so.

\begin{acknowledgments}
\noindent We wish to thank Tiziano Camporesi, Joel Butler, and the CMS collaboration for their encouragement. We would also like to thank Vladimir Ivanchenko, Andrea Dotti and Mihaly Novak for useful discussions regarding {\sc Geant4}.
\end{acknowledgments}
\bibliography{milliQan_LOI}

\end{document}

%% file: intro.tex
\noindent In an effort to explain galactic dark matter, the idea of additional ``sectors" beyond the SM has received considerable attention with a variety of examples (hidden valleys, secluded sectors, dark sectors, hidden sectors). The experimental searches for evidence of these sectors are driven largely by two factors:  the particular way by which the extra sector is coupled to the SM; and the mass scale(s) in the extra sector. The specific details of the extra sector (the precise gauge group, the number of extra particles, etc.) are often not of great importance in designing the search. 

In this LOI we propose a dedicated experiment to search for milli-charged particles (mCP). Since the search depends only on the mass and charge of such particles, one can view this search as part of the general program to search for additional sectors but to give a concrete example, we consider an extra abelian gauge field that couples to a massive Dirac fermion (``dark QED") and that mixes with hypercharge through the kinetic term~\cite{Holdom:1985ag},
\be
\mathcal{L} = \mathcal{L}_{\rm SM} &+& \mathcal{L}_{\rm extra-sector} \\ \nonumber
~\\ \nonumber
\mathcal{L}_{\rm extra-sector} &=&  -\frac{1}{4}A'_{\mu\nu}A^{\prime\mu\nu} + i\bar{\psi}\left( \slashed{\partial}+ie' \slashed{A}' +i \MQ\right)\psi \\ &~& -\frac{\kappa}{2} A'_{\mu\nu}B^{\mu\nu} .\
\label{eqn:def_lag}
\ee
Here $\psi$ is a Dirac particle of mass $\MQ$ that is charged under the new $U(1)$ field $A'_{\mu}$ with charge $e'$, and the field-strength is defined as $A'_{\mu\nu} = \partial_\mu A'_\nu - \partial_\nu A'_\mu$. The last term in Eq.~(\ref{eqn:def_lag}) is a kinetic mixing term between the field strength of the new gauge boson and that of hypercharge~\footnote{Such a term is expected in grand unified theories and more generally whenever there exists massive fields that are charged under both hypercharge and the new gauge boson, even when these heavy fields are not accessible at low energies. }.

Eliminating the mixing term by redefining the new gauge boson as, $A'_\mu \rightarrow A'_\mu + \kappa B_\mu$  results in a coupling of the charged matter field $\psi$ to hypercharge (as well as an immaterial redefinition of the hypercharge coupling), 
\be
\nonumber
\mathcal{L} &=& \mathcal{L}_{\rm SM}  -\frac{1}{4}A'_{\mu\nu}A^{\prime\mu\nu} \\ &+& i\bar{\psi}\left( \slashed{\partial}+ie' \slashed{A}' - i \kappa e' \slashed{B} + i \MQ \right)\psi .\
\label{eqn:mCP_lag}
\ee
The new matter field $\psi$ therefore acts as a field charged under hypercharge with a charge $\kappa e'$, a milli-charge~\cite{Holdom:1985ag}. The mCP $\psi$ couples to the photon and $\ZZ$ boson with a charge $\kappa e'\cos\thetaW $ and $-\kappa e'\sin\thetaW $, respectively. The fractional charge in units of the electric charge is therefore $\epsilon \equiv \kappa e' \cos\thetaW/e$, where $\epsilon \ll 1$. 

Previous experiments have looked for non-quantized charged particles~\cite{Prinz:1998ua, Davidson:2000hf, Badertscher:2006fm, CMS:2012xi}. The parameter space spanned by the mass and charge of the mCPs is also constrained by indirect observations from astrophysical systems \cite{Davidson:1991si, Mohapatra:1990vq, Davidson:1993sj, Davidson:2000hf}, the cosmic microwave background~\cite{Dubovsky:2003yn, Dolgov:2013una}, big-bang nucleosynthesis \cite{Vogel:2013raa}, and universe over-closure bounds~\cite{Davidson:1991si}. While direct searches robustly constrain the parameter space of mCPs, indirect observations can be evaded by adding extra degrees of freedom. In particular, the parameter space for mCPs with masses $\MQ$ $0.1 \lsim\MQ\lsim100\GeV$ is largely unexplored by direct searches.

A natural question is whether the current general-purpose LHC experiments can improve sensitivity in the parameter space for mCPs with masses $\MQ$ $0.1 \lsim\MQ\lsim100\GeV$.  A recent analysis looking for low ionizing particles in CMS excluded particles with charge $\pm e/3$ for $\MQ < 140\GeV$ and particles with charge $\pm2e/3$ for $\MQ < 310\GeV$~\cite{CMS:2012xi}. For fractional charges much smaller than unity, mCPs could be detected in monojet plus missing energy searches. However, for electroweak production of new quasi-invisible states, such as the mCPs studied here, the penalty on the cross-section associated with initial state radiation precludes CMS/ATLAS sensitivity to such particles, even with the very large datasets envisaged for the HL-LHC. Thus, to detect mCPs at the LHC, an alternative experimental strategy is needed. 

In Ref.~\cite{Haas:2014dda} some of the authors of this LOI proposed a new search to be conducted at the LHC with a dedicated detector targeting this unexplored part of parameter space, namely mCP masses $0.1 \lsim \MQ \lsim 100 \GeV$, for charges $Q$ at the $10^{-3}~e -10^{-1}~e$ level. The experimental apparatus envisaged in this paper was one or more scintillator detector layers of roughly 1 m$^3$ each, positioned near one of the high-luminosity interaction points of the LHC. The experimental signature would consist of a few photo-electrons (PE) arising from the small ionization produced by the mCPs that travel unimpeded through material after escaping the ATLAS or CMS detectors. The proposed experiment is a model-independent probe of mCPs, since it relies only on the production and detection of mCPs through their QED interactions. 

We have found the PX56 Observation and Drainage gallery above the CMS underground experimental cavern (UXC) to be an ideal site for such an experiment, and we propose in the following to situate there our experimental apparatus, which we call milliQan.

%% file: site.tex
\noindent LHC Point 5, Cessy is host to the Compact Muon Solenoid (CMS) experiment and its supporting infrastructure. As such, the appropriate services are available for the installation and operation of milliQan. We propose that the detector be set in the PX56 drainage gallery located above CMS UXC. Material access is limited by a door measuring $0.9~\mathrm{m}\times2.1~\mathrm{m}$. That door links the drainage gallery and the platform installed in the shaft. This platform is within the coverage area of the overhead crane installed in the assembly hall building (SX5).  Any components larger than a regular toolbox will have to be lowered to the gallery using the overhead crane and then passed through this door. Personnel access will be through a sector door located between the PM54 staircase and the drainage gallery. The door is interlocked with the LHC access control system. This entails that no access to milliQan will be possible while the LHC is in operation. The electrical infrastructure available in the CMS service cavern (USC55) can be expanded to bring power to milliQan using existing cable channels in the PM54 shaft. Our studies indicate that forced air from a portable air conditioning unit should provide sufficient cooling to hold the PMTs somewhat below room temperature and reduce the backgrounds from dark noise.

\begin{figure}[htp]
\centering
\includegraphics[width=0.99\linewidth]{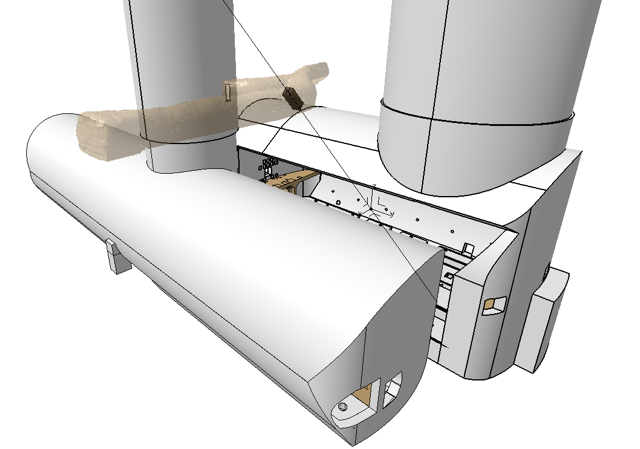}
\caption{3D model showing optimal position of milliQan within the PX56 Drainage and Observation gallery located above CMS UXC.\label{fig:site}}
\end{figure}

The proposed gallery is limited in space. Knowing this, a 3D model was combined with a laser scan of the gallery to give a best as-built estimation. Using the model an optimal position was found which offered; a distance to the IP of 33~m, 17~m of which is through rock, an angle of 43.1 degrees from the horizontal plane. Under these conditions clearance between the corners of the detector and the gallery can be as little as 30~mm. The selected location in the 3D model is shown in Figure~\ref{fig:site}. Clearance increases as the detector is moved away from the IP so the final position may come down to a compromise. These values are current best estimates; some small amounts of variation in the final design should be expected.

While the installation of the detector is not anticipated to present major challenges, a detailed work sequence is being studied. It will allow identifying the logistics operations required to move the components to their final destination in the drainage gallery and the corresponding necessary tooling. The study will also investigate a workflow for assembly, mounting and then alignment of the detector. A preliminary assessment shows that the CERN survey group will be able to align milliQan with a precision of about 2~cm. Safety aspects associated with installation, operation and maintenance of the detector will also have to be further studied. Fire loads and access requirements are all constraints to be considered. It is already certain that the detector support structure must allow a passage of at least 60~cm through the gallery. During access periods, this will require the detector to be moved into a ``stowed" position where the length of the detector will be parallel with the wall of the gallery. 

In terms of beneficial occupancy of the drainage gallery, this proposal has already been run through various LHC operation bodies and no conflicts were identified. In particular, we have received confirmation from the LHC integration office that they have no plans to ever install anything in the drainage gallery and that from their point of view it is available for use for milliQan.  While still a lot of design and integration work has to be done, it seems that this scheme has a strong chance of meeting the requirements to install milliQan at point 5 in a reasonably short time.

%% file: cms.tex
\noindent A design requirement is that milliQan would be operated with no interference to CMS operations. The milliQan detector will self-trigger to a dedicated readout as described in Section~\ref{sec:daq}. The only critical needs from CMS would be basic infrastructure such as power and ethernet, and read access to the CMS BRIL DB in order to receive information on the luminosity delivered to Point 5. The experiment would also need access to the LHC clock, which could be provided by a TCDS fiber. TCDS fibers can also provide run/luminosity section/orbit markers that would be used to synchronize milliQan with the CMS luminosity system. Finally, while not a requirement, access to the BPTX signals via LEMO cables would be useful to be able to distinguish collisions from empty buckets.

%% file: det.tex
\noindent The detector that we propose to install at the location described in Section~\ref{sec:site} is a~$1~\mathrm{m}\times1~\mathrm{m}\times3~\mathrm{m}$ plastic scintillator array. The array will be oriented such that the long axis points at the nominal CMS IP. The array is subdivided into 3 sections each containing 400~$5~\mathrm{cm}\times5~\mathrm{cm}\times80~\mathrm{cm}$ scintillator bars optically coupled to high-gain PMTs. A triple-incidence within a 15 ns time window along longitudinally contiguous bars in each of the 3 sections will be required in order to reduce the dark-current noise, which we expect to be the dominant background~\footnote{The detector will be adequately shielded from other background sources such as activity in the scintillator and environmental radiation.}.

A $Q=1e$ minimum-ionizing charged particle leaves roughly 2~MeV/cm in a material of density 1 g/cm$^3$~\cite{Beringer:1900zz}. For plastic scintillator, such energy deposition results in about $10^4$ photons per MeV, meaning $1.6\times10^6$ photons would be liberated in a 80~cm long scintillator. For a particle with electric charge $Q<1e$, the energy deposition is reduced by the factor of $Q^2$ resulting, for fractional charges at the lower limit of the sensitive range, in just a few photons liberated on average in the same 80~cm long scintillator. Allowing for an estimated detection efficiency of about 10\%, we therefore expect an average of $\mathcal{O}(1)$ photoelectron (PE) from each attached phototube for each mCP with $Q=\mathcal{O}(10^{-3})~e$ that traverses our 80~cm plastic scintillator~\cite{Prinz:1998ua}. The signal we will search for is a longitudinal triple-incidence of hits with one or more PEs. Requiring triple-incidence will control the background to $\mathcal{O}(10)$ events per year with $N_{\rm PE}\ge 1$, as discussed in Section~\ref{sec:bkg}.

%% file: infra.tex
\noindent We are developing the mechanics to support the detector in the drainage gallery site so as to allow both modular assembly and movement of the detector to a stowed position during access periods.

\begin{figure}[htp]
\centering
\includegraphics[width=0.8\linewidth]{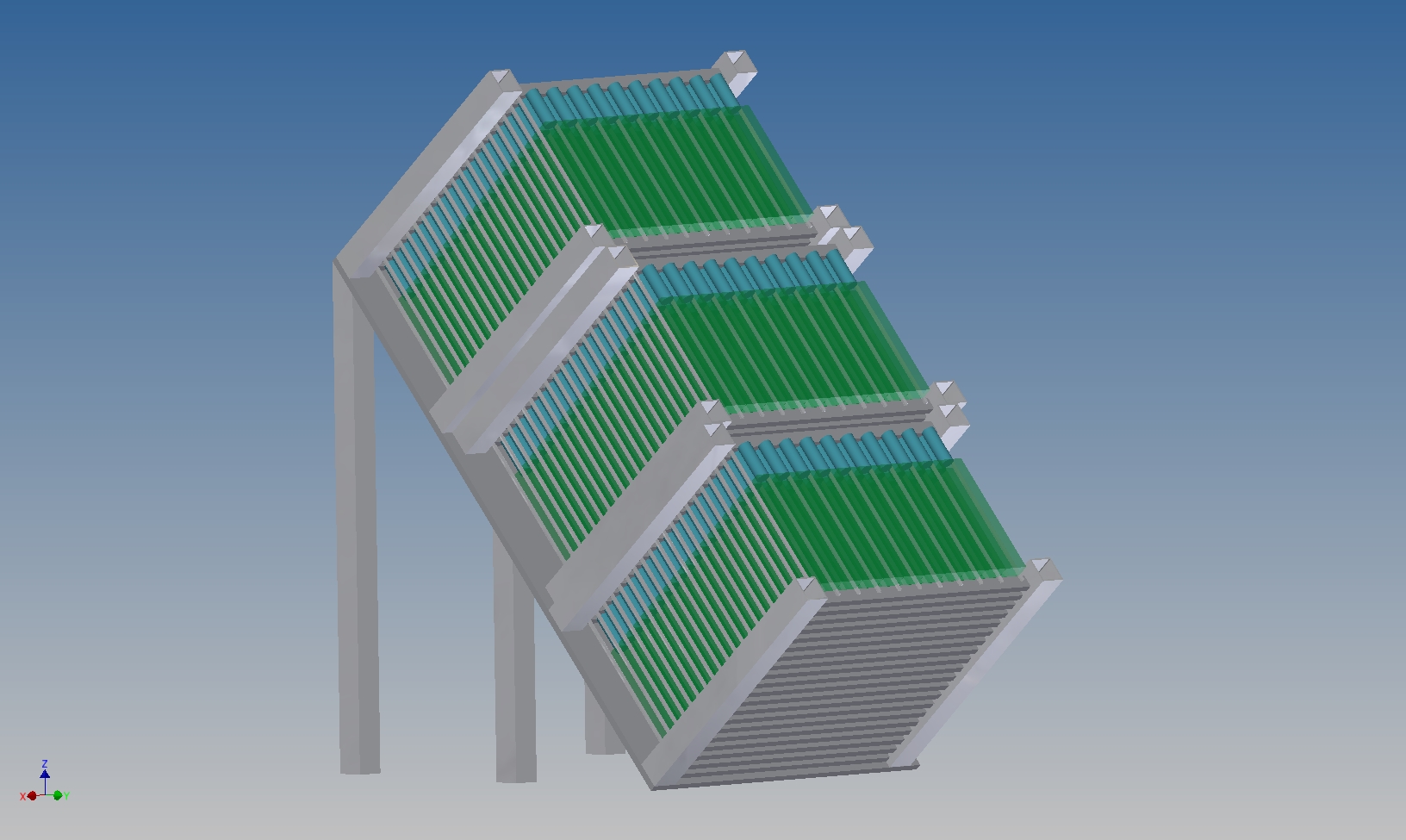}
\caption{A sketch of the working design for the mechanical support. The rows of scintillator bars and PMTs are mounted in trays within three separate stacks.
The middle stack is offset horizontally by 1/2 unit as discussed later in this document. An adjustable platform supports all three stacks and can be tilted to point toward the collision point for data taking or be retracted to a horizontal orientation during access periods. 
\label{fig:mechsketch}}
\end{figure}

We have a working design consisting of three stacks, one for each of the three layers along the mCP flight direction, that can be separately assembled.
The scintillator bars are mounted on trays within each stack as sketched in Fig.~\ref{fig:mechsketch}. The three stacks, each of which might weigh up to $\sim2000$~kg, will be mounted on an adjustable platform that can tilt the full assembly to point toward the collision point, or be retracted to a horizontal orientation to be moved aside during access periods.

The sides of the assembly would be covered with a mu-metal skin for magnetic shielding and enclosed for temperature control. The size of the drainage gallery leaves little clearance at the corners of the detector, which constrains the thickness and mounting in those regions. However, the space along the sides of the detector is not constrained. We envision using that space for the hoists necessary for positioning and retractions as well as the cable plant and cooling.

As discussed below, the dominant background is expected to arise from PMT dark current, which can be reduced by cooling the PMTs below room temperature. The limited infrastructure available in the drainage gallery hinders the use of water cooling, but we expect the estimated heat load of a few kW could be managed with forced air cooling from a locally positioned air conditioning unit. 

Such a design appears to be workable without significant infrastructure demands.  We are proceeding to study specific engineering options in more detail.

%% file: calib.tex
\noindent PMTs that meet the required specifications in terms of pulse rise time, dark current and counting rates, and quantum efficiency require applied high voltages (HV) between $1-2.5$~kV and have maximum current ratings of $0.2-0.5$~mA. For the detector design using 12 read-out channels per module, the HV power supply (HVPS) must provide approximately 10~mA per module. In order to minimize costs, we aim to use one HV power supply to power 10 modules (120 channels), and thus we require a HVPS rated to approximately 100~mA and 500~W. For a 100 module detector, 10 HVPS are required and the total power requirement would thus be approximately 5~kW. Several commercial HVPS systems exist that meet these requirements. For example, the \href{http://theelectrostore.com/shopsite_sc/store/html/PsslashEK03R200-GK6-Glassman-New-refurb.html}{Glassman model number PS/EK03R200-GK6} provides an output of $\pm3$~kV with a maximum of 200 mA, and features controllable constant current / constant voltage operation. Regulation and monitoring of the power supplied to the detector will be required on both the module distribution boards and the front-end distribution boards. In both cases, over-current and over-voltage protection will be necessary both for safety and in order to protect the front-end electronics from damage. The monitoring may be accomplished by a measurement circuit that digitizes and transmits the measured voltages and currents over a serial bus to the slow control system for the detector by a generic, CERN built data acquisition board called an Embedded Local Monitoring Board (ELMB)~\cite{ELMB}.

Energy calibration will be done in situ using an $^{241}$Am source, which yields a 60~keV $X$-ray. Calibration runs performed at specified intervals will track the PMT+scintillator response as a function of time. In addition to energy calibration, an LED pulser that can deliver a stable light pulse into each scintillator will also be deployed. The LED system will be used to monitor drift in response of the PMT+scintillator as a function of time in between $^{241}$Am source calibrations as well as detect any inefficient or non-functional readout channels.

%% file: readout.tex
\noindent Analog pulses from the PMTs must be read out, digitized, and stored for offline analysis. Furthermore, as the pulse rate per PMT is large, a trigger will be used to record only those pulses during interesting time windows when significant activity in the detector is observed, with at least two of the three layers in a $2\times2\times3$ detector module having a pulse above the single PE threshold. Fortunately, commercial electronics are available for performing these tasks in a simple way at reasonable cost. 

The triggering and readout can be performed by the CAEN V1743 digitizer, which uses the SAMLONG chip. The VME board has 16 channels, each of which is sampled at 3.2 GS/s (a sample each 312.5 ps) into a 1024 analog buffer ring (320 ns long). Analog noise is about 0.75 mV per channel, allowing good identification of and triggering on single PE signals, which can easily be above 3 mV. The only drawback of the analog buffer approach is that the board is dead while digitizing the buffer, which takes 20-125 $\mu$s, depending on how much of the buffer, i.e. what fraction of the 320 ns, is digitized. This results in a deadtime of $\lsim10~\%$, even at 1kHz trigger rate, and the time within $\sim10~\mu$s after a trigger is anyways of low quality due to the presence of PMT afterpulses. Seven buffers are available to hold triggered events in RAM until it is read out, either over VME or an optical link. Thus there is no dead time from readout, up to rates of $\sim1$~kHz. The digitization is done with 12 bits of precision, and the dynamic range is 2.5 V, allowing good resolution of 2500 mV / 4096 = 0.61 mV. 

The board is also capable of self-triggering on each channel's analog sample, via a 16-bit discriminator with a resolution of 0.04 mV, for each channel. The triggers for each pair of channels (0+1, 2+3, etc.) are then ORed (or ANDed) together within the board FPGA, so that 8 trigger bits are available per event. These are then compared, further in the board's FPGA, and can be required to form coincidences within an adjustable length time window. We envision ORing together the pairs of channels, which would be connected to channels in the same longitudinal $2\times2$ detector layer. We then would ask that at least 2 such bits are fired within a 15 ns window (the smallest time window available, but roughly optimal for our pulse resolution and triggering needs). This means that at least two scintillator bars (not in the same part of a layer) are above threshold coincidently. When a coincidence is triggered in one board's channels as just discussed, it can be propagated to the other boards of the experiment via an external trigger lemo input. Data will be read out via CAEN CONET 2 over an optical fiber to a PCI card, which can sustain up to 80 Mbps, which is an order of magnitude more than the expected event data rate, as discussed in section \ref{sec:bkg}.

The CAEN V1743 digitizer board is capable of handling an external clock, and distributing it across multiple boards via an external clock sync in lemo cable. And the board has a 16-bit LVDS input that can be used as an event identifier to tag events for offline identification. Additional background rejection can be obtained by requiring scintillation signals to be coincident with the expected arrival time of particles produced in LHC collisions. This can be achieved by sending the LHC clock to the readout system and recording the phase of the clock for each triggered event. The clock signal will be sent on a fiber from the CMS Trigger and Clock Distribution System (TCDS), which will also carry CMS run start and stop signals, as well as ``luminosity section'' and orbit counter reset signals. These signals will allow recorded events to be correlated with the CMS luminosity database, enabling the cross-section of a signal to be measured. The TCDS fiber will be received by a dedicated clock and timing module, comprising an FPGA, an SFP carrier for reception of the optical fiber, and LVDS outputs to the readout cards. (Various FPGA evaluation boards are suitable, for example the Xilinx Spartan 6 SP605). The TCDS signals will be decoded in the FPGA, which will contain counters recording the LHC clock phase, as well as the CMS run, luminosity section and orbit number, for each trigger.

%% file: bkg.tex
\noindent The dominant background is expected to come from dark-current pulses in the PMTs. Additional, sub-dominant, sources of background include activity in the scintillator from background radiation and photo-multiplier afterpulsing. In Ref.~\cite{Haas:2014dda} we assumed a dark-current background rate for a single PMT to be 550 Hz for a single-PE threshold. To get a more accurate estimation of this rate, we constructed a test setup using a 3" Bicron-412 scintillator coupled to a 3" Hamamatsu H2431 PMT at 3 kV, readout with a CAEN V1743 digitizer, described in Section~\ref{sec:daq}. Once the board receives a triggering event, the analog buffer is digitized. During this digitization window, which lasts approximately 100 $\mu$s, no additional trigger can be accepted. Using this test setup, we measured a dark-current rate of approximately 1~kHz at room temperature. By reducing the high voltage and cooling the PMT we are able to significantly reduce the background rate.  The studies are still ongoing, but with these handles and by optimizing the choice of PMT we expect to be able to bring the rate to below 500~Hz.
 
The vast majority of pulses from background radiation, including cosmic muons, will consist of a large number ($>$1000) of PEs. By implementing an offline veto of these large pulses, for example those with more than 10 PEs, these backgrounds can be dramatically reduced. Whenever a pulse enters a photo-multiplier, there are smaller after-pulses that are generated. These small after-pulses occur within approximately 10~$\mu$s of the original large pulse. Since the large pulse will trigger the board, the small after-pulses will fall within the digitization deadtime of the CAEN V1743 and thus be vetoed.
  
The rates from all background sources will be greatly reduced by the requirement of a 3-fold coincidence between the layers of the detector.  However, a remaining background source is cosmic muons which pass along the edge of all three scintillator layers. Such a glancing trajectory could result in a small pulse in each of the 3 PMTs.  To account for this possibility, we will offset the middle layer, eliminating such trajectories. The remaining background from cosmic muons would be from trajectories that glance by the first and third layers and pass through the bulk of the second layer. 


With the dominant background coming from the dark-current in the PMT, we assume a total background rate per PMT of $\nu_B = $500 Hz. By reading out pairs of PMTs in the same layer, the rate per group would be $2\nu_B$. Each CAEN V1743 board would receive input from 6 such pairs, 2 in each layer, leaving 2 groups without input. With a reasonably robust time window of $(\Delta t)_{\mathrm{online}} =100$~ns, the double coincident trigger rate per-board will be $\binom{6}{2}\left(2\nu_B\right)^2(\Delta t)_{\mathrm{online}} =1.5$~Hz. The entire detector will be read out if one board triggers and there will be 50 such boards in total. Therefore, the full background trigger rate is expected to be 75~Hz. Offline we expect to be able to tighten the time window to $(\Delta t)_{\mathrm{offline}} =15$~ns as discussed previously. The offline background rate for a triple coincidence is given by $\nu^3_B(\Delta t)_{\mathrm{offline}}^2 = 2.8\times10^{-8}$~Hz. Since there are 400 such sets, the total offline background rate is estimated to be $1.1\times10^{-5}$~Hz. By the end of Run 3 LHC will have delivered 300 fb$^{-1}$. Assuming an average instantaneous luminosity of $2\times10^{34}$~cm$^{-2}$s$^{-1}$ and a 50\% timing window efficiency, we calculate a trigger live-time of $1.5\times10^{7}$~s. During HL-LHC operation, 3000 fb$^{-1}$ will be delivered with an instantaneous luminosity $1\times10^{35}$~cm$^{-2}$s$^{-1}$, corresponding to a a trigger live-time of $3.0\times10^{7}$~s. We therefore estimate that we will have 165 (330) background events in 300 (3000)~fb$^{-1}$. However, we expect to be able to use additional handles (e.g. pulse shapes of the pulses and tighter timing cuts) to further reduce the total background in these run periods to $\sim50$ (100) events.


These studies provide an estimate of the background rate, which is used below to calculate the expected sensitivity. Ultimately, the background will be determined from the data, for example by measuring the rate of near signal-like events such as triple-incidence of single PE hits that do not point to the collision.

%% file: sim.tex
\begin{figure}[htp]
\centering
\includegraphics[width=0.99\linewidth]{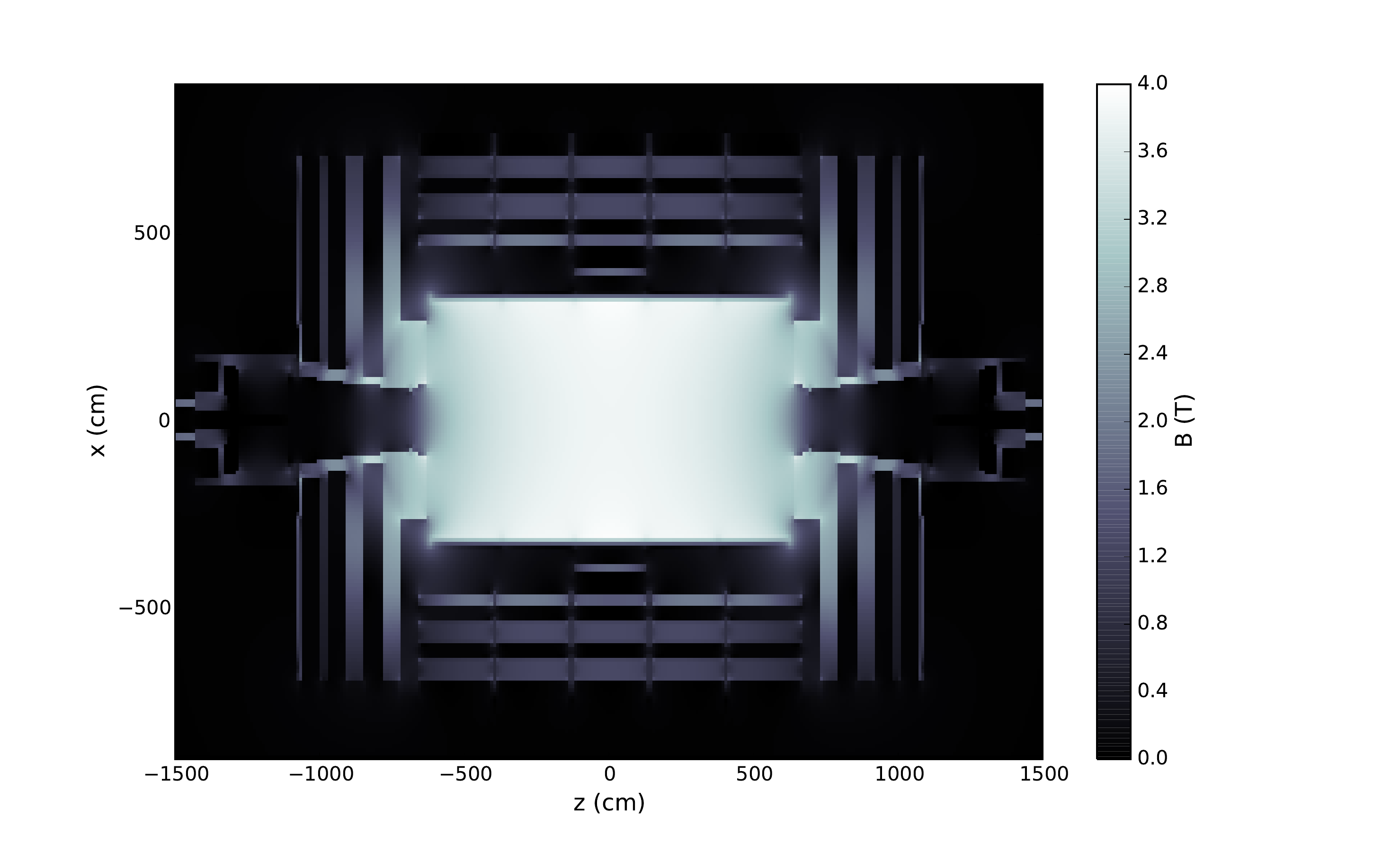}
\caption{Map of the CMS magnetic field in the r--z plane.\label{fig:bfield}}
\end{figure}
\begin{figure}[htp]
\centering
\includegraphics[width=0.8\linewidth]{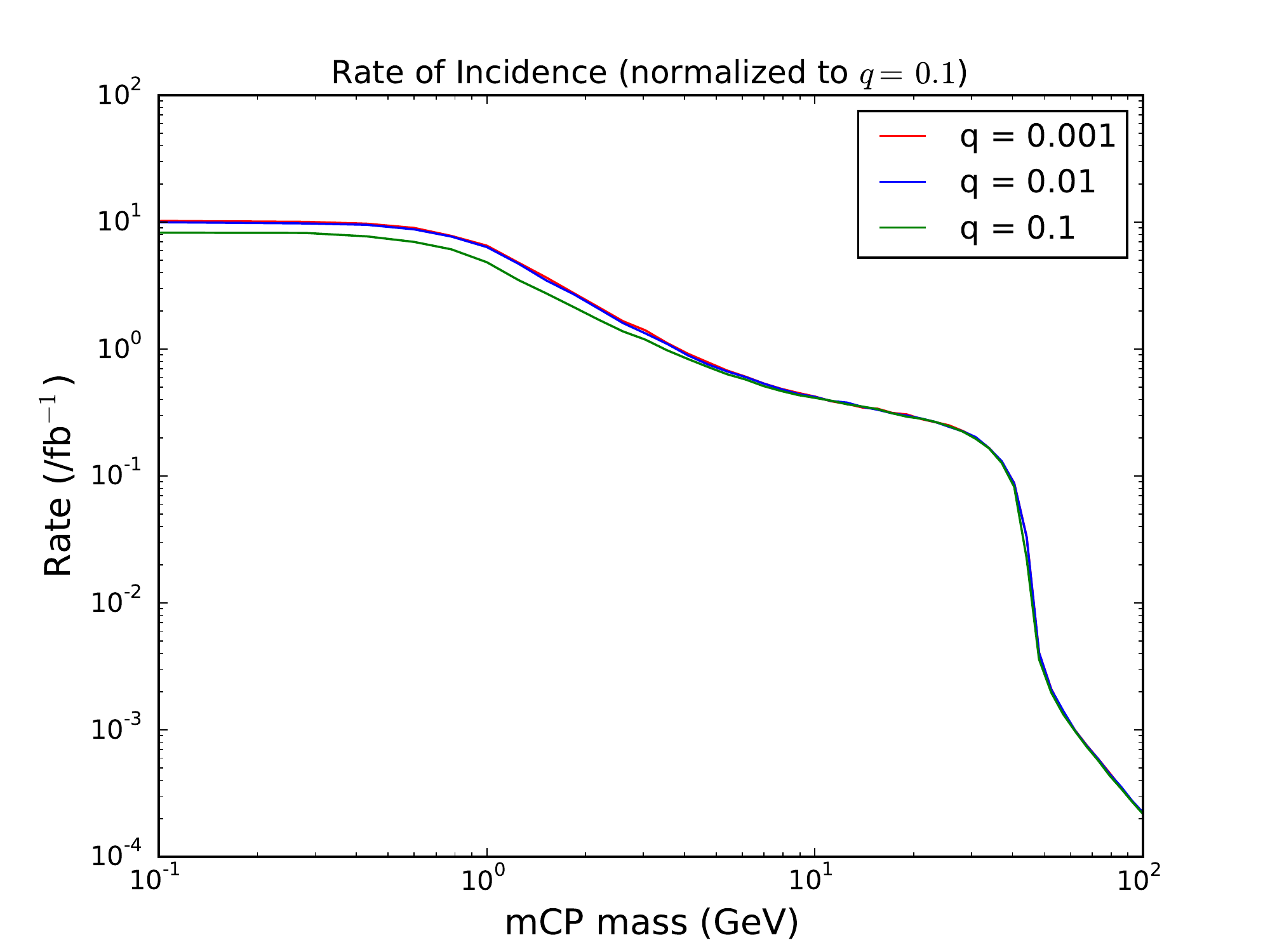}
\caption{Number of expected mCP particles per fb$^{-1}$ of integrated luminosity incident at the detector as a function of the mass of the milli-charged particle. To illustrate the dependence of the acceptance on the charge, the $Q^2$ production dependence has been factored out by normalizing the cross section for all charge scenarios to that for a milli-charged particle with $Q=0.1~e$. \label{fig:mcpRate}}
\end{figure}

\noindent In order to evaluate the projected sensitivity of the experiment for various mCP electric charges and masses, we performed a full simulation of the experiment, including a {\sc Geant4}~\cite{Agostinelli:2002hh} model of the detector. The simulation is performed in two stages.  In the first, {\sc Feynrules}, {\sc Madgraph5} and {\sc Madonia} were used to simulate the production of mCP particles via Drell-Yan, J/$\Psi$, $\Upsilon$(1S), $\Upsilon$(2S), and $\Upsilon$(3S) channels at 14 TeV center-of-mass energy~\cite{Alwall:2014hca,Alloul:2013bka,madonia}. Particles produced at the interaction point were propagated using a map of the CMS magnetic field, shown in Figure~\ref{fig:bfield}, to the proposed experimental site described in Section~\ref{sec:site}. Although small for particles with charge $Q \ll e$, the effects of multiple scattering and energy loss were included using a simplified model of the CMS detector material budget and a region of rock spanning 17~m between the CMS experimental cavern and the proposed experimental site. The number of expected mCP particles per fb$^{-1}$ of integrated luminosity incident at the detector is shown in Figure~\ref{fig:mcpRate} as a function of the mass of the milli-charged particle.  To illustrate the dependence of the acceptance on the charge, the cross section for all charge scenarios is normalized to that for a milli-charged particle with $Q=0.1~e$.

\begin{figure}[htp]
 	\centering
	 \subfloat[]{\includegraphics[width=0.8\textwidth]{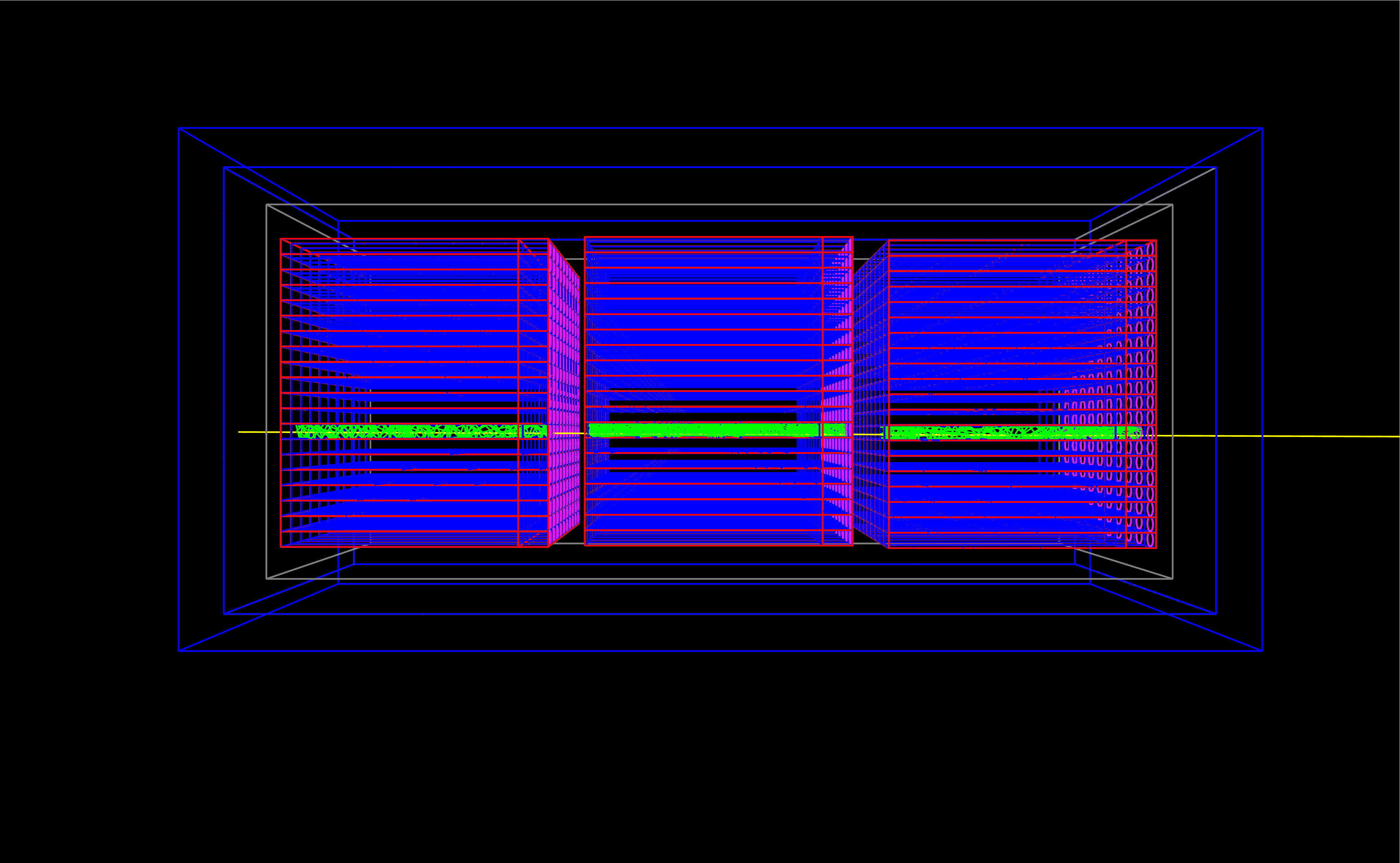}\label{fig:DetectorSideActive}}\quad
	\subfloat[]	{\includegraphics[width=0.8\textwidth]{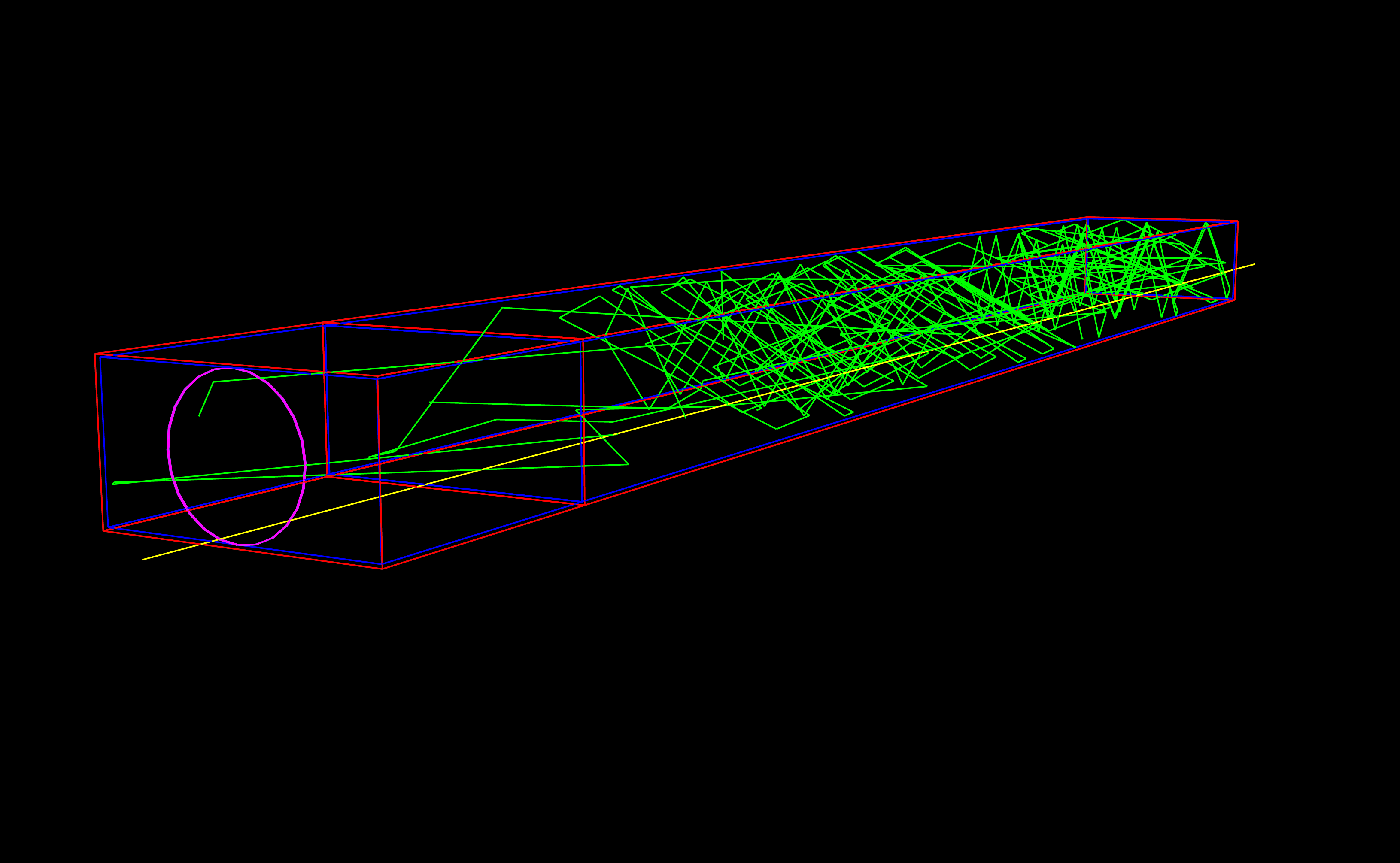}\label{fig:BlockDiagonalFrontActive}}%
	\caption{Depiction of the \protect\subref{fig:DetectorSideActive} full detector and \protect\subref{fig:BlockDiagonalFrontActive} a single scintillating block with coupled phototube, as implemented in the {\sc Geant4} detector simulation. The mCP is yellow and radiated photons are green. \label{fig:detGeant}}
 \end{figure}

\begin{figure}[htp]
 	\centering
 	\subfloat[]{\includegraphics[width=0.8\textwidth]{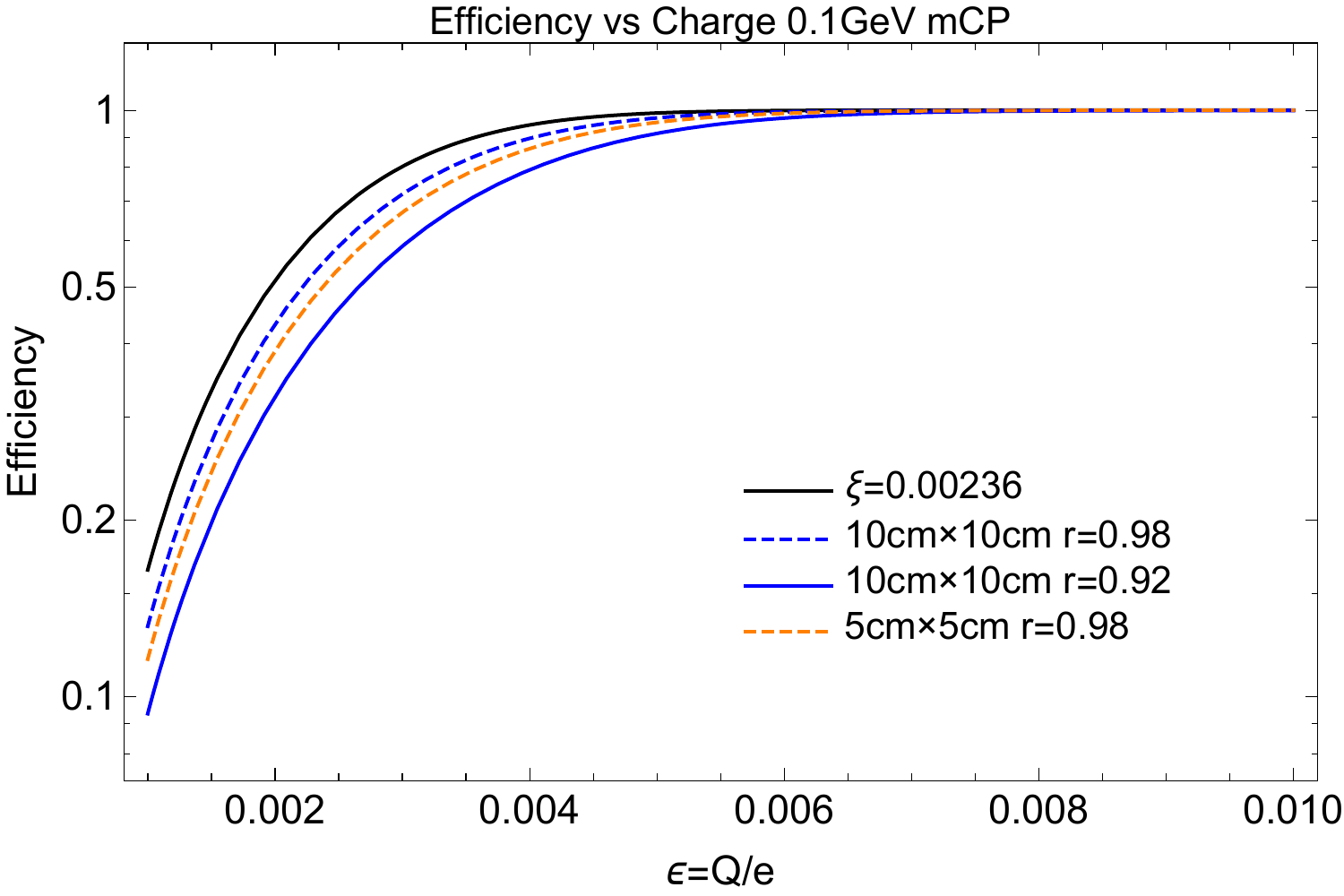}\label{fig:deteffsingle}}\quad
         \subfloat[]{\includegraphics[width=0.8\textwidth]{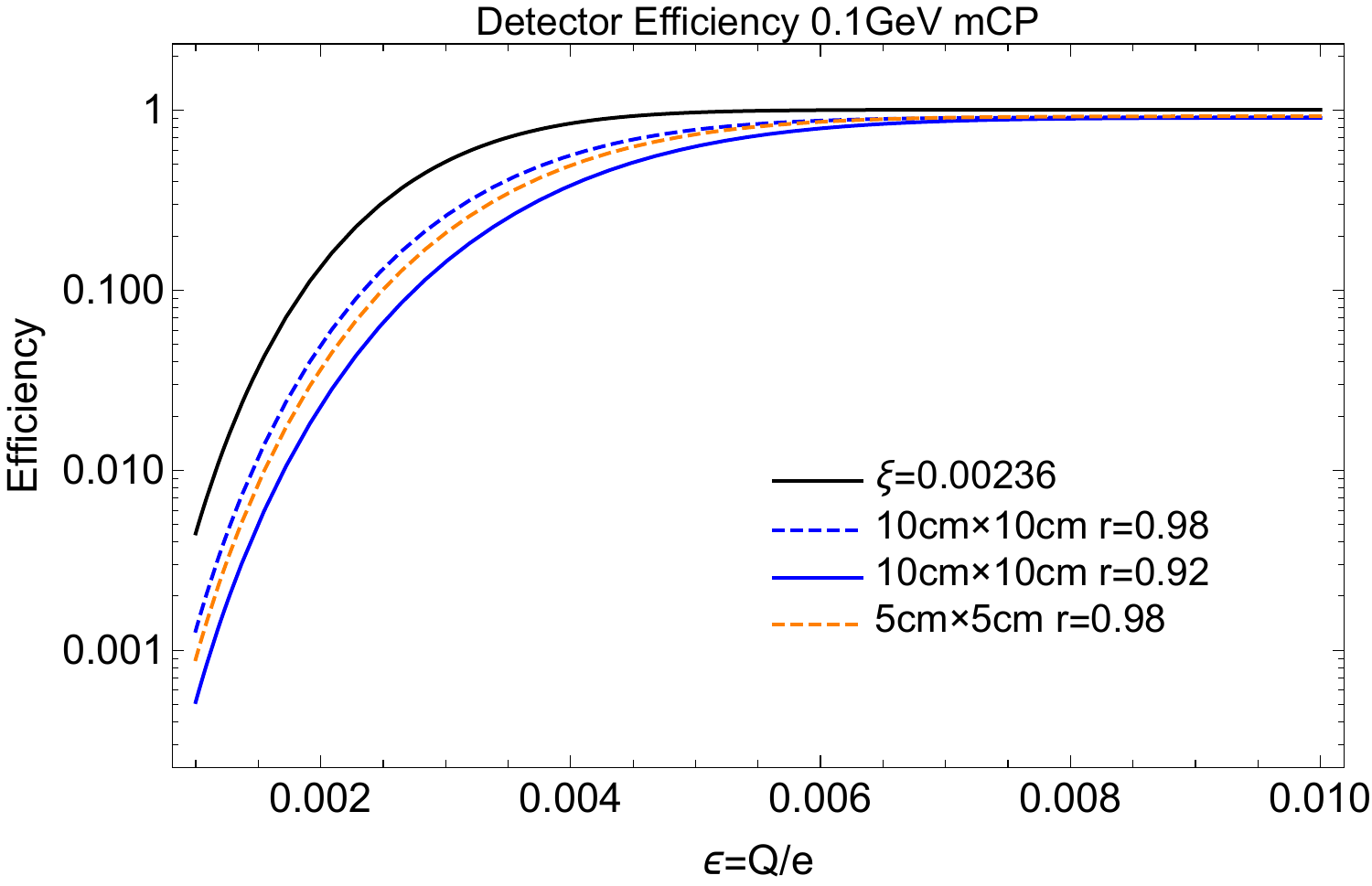}\label{fig:deteffall}}%
     \caption{Efficiencies for \protect\subref{fig:deteffsingle} a single scintillator block and coupled PMT and \protect\subref{fig:deteffall} the whole detector with 15ns triple-incidence, as determined from the {\sc Geant4} detector simulation. \label{fig:detEff}}
 \end{figure}

In the second stage, we calculated the signal efficiency by running the kinematic distributions of the particles at the proposed experimental site through a full {\sc Geant4} simulation of the detector, as viewed in Figure~\ref{fig:detGeant}. This was important because although we can estimate part of the efficiency of the detector by hand, the small charge regime is sensitive to details such as the reflectivity, the light attenuation length, and the shape of the scintillator. These details, as well as the quantum efficiency, light emission spectrum and the fast time constants are modeled in {\sc Geant4} using the specifications provided by the manufacturers for Saint-Gobain BC-408 plastic scintillator and Hamamatsu R329-02 PMTs~\cite{ScintBC,PMTHamamatsu}. We defined a new fermion of variable mass and electric charge. Its electromagnetic interactions consist of multiple scattering, Bethe-Bloch energy loss and density effects, implemented in {\sc Geant4} using the ``G4WentzelVIModel'' and ``G4hIonisation'' packages, which are documented in the source files. Figure~\ref{fig:detEff}\subref{fig:deteffsingle} shows how the efficiency of a single scintillator bar changes as a function of electric charge when varying the transverse dimensions and the reflectivity, for a 0.1~GeV mCP. Figure~\ref{fig:detEff}\subref{fig:deteffall} shows the same plot for the full detector, requiring a 15~ns triple coincidence. In both plots, we compare the {\sc Geant4} efficiencies to the efficiencies assumed in the Ref.~\cite{Haas:2014dda}. The probability of seeing one or more photoelectrons in each layer of the detector was parametrized using Poisson statistics by
\be
P=\left(1-\exp\left[-N_{PE}\right]\right)^3,
\label{eq:efficiency}
\ee
where $N_{PE}=\left(\frac{Q}{\xi}\right)^2$ is the average number of photoelectrons produced for a given charge. The constant of proportionality $\xi$ was estimated by finding the electric charge that gives 1 photoelectron, given the material light yield, a 10\% detection efficiency, the length of the scintillator and typical energy deposits of a minimally ionizing particle. It was found to be $\xi \approx 0.0024$. Comparing this estimate to the {\sc Geant4} efficiencies, we find good agreement, especially for the large mass regime (not shown). The mCPs in the lower mass regime are more relativistic, and deposit less energy. \\

\begin{figure}
   \centering
   \includegraphics[width=0.8\linewidth]{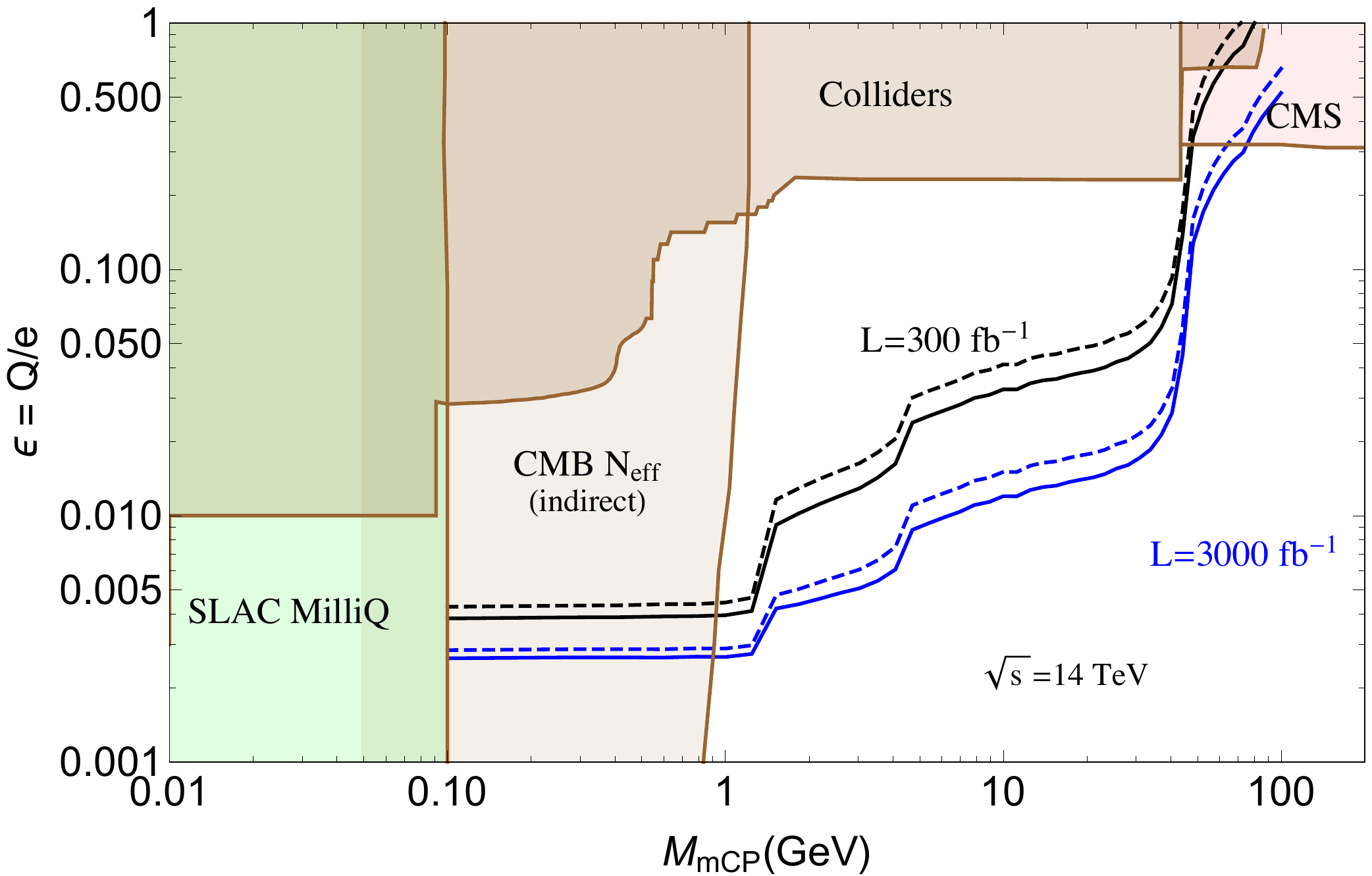}
   \caption{Expected sensitivity for different LHC luminosity scenarios. The black line shows the expected 95\% C.L. exclusion (solid) and $3\sigma$ sensitivity (dashed),
assuming $300~\text{fb}^{-1}$ of integrated luminosity. In blue we show the corresponding expectations for $3000~\text{fb}^{-1}$.
\label{fig:abc}}
\end{figure}

Combining the estimated background rates discussed in Section~\ref{sec:bkg} with the cross-sections, acceptances and efficiencies calculated for all masses and electric charges, the sensitivity projections of the milliQan experiment for LHC and HL-LHC are shown in Figure~\ref{fig:abc}. 

